\documentclass[11pt]{article}
\usepackage[portuges,brazilian,english]{babel}

\usepackage{amssymb}
\usepackage{graphicx}
\usepackage[T1]{fontenc}
\begin{document}

\title{Ruling out the Modified Chaplygin Gas Cosmologies}
\author{J. C. Fabris\footnote{e-mail: fabris@pq.cnpq.br}, H.E.S. Velten\footnote{e-mail: velten@cce.ufes.br} \\
Departamento de F\'{\i}sica, Universidade Federal do Esp\'{\i}rito Santo, \\
CEP 29060-900 Vit\'{o}ria, Esp\'{\i}rito Santo, Brasil\\
\\
C. Ogouyandjou\footnote{e-mail: ogouyandjou@imsp-uac.org} and J. Tossa\footnote{e-mail: joel.tossa@imsp-uac.org}\\
Institut de Math\'ematiques et de Sciences Physiques - IMSP\\
B.P. 613, Porto Novo, B\'enin
}
\date{}
\maketitle

\begin{abstract}
The Modified Chaplygin Gas (MCG) model belongs to the class of a unified models of dark energy (DE) and dark
matter (DM). It is characterized by an equation of state (EoS) $p_c = B\rho - A/\rho^{\alpha}$, where the case $B=0$ corresponds to the Generalized Chaplygin Gas (GCG) model. Using a perturbative analysis and power spectrum observational data we show that the MCG model is not a sucessful candidate for the cosmic medium unless $B=0$. In this case, it reduces to the usual GCG model.
\end{abstract}

\vspace{0.5cm} \leftline{PACS: 98.80.-k, 04.62.+v}

\section{Introduction}

The cross of different observational results at cosmological level indicates that besides the usual expected contents of the cosmic budget, like
baryons and radiation, there is a dark sector, with two components, dark matter and dark energy.
In principle, dark matter is present in local structures like galaxies and cluster of galaxies, suffering consequently the process of gravitational
collapse. In this sense, dark matter behaves much as like ordinary matter. However, it does emit any kind of eletromagnetic radiation. Dark energy, on the other hand, seems to remain a smooth, not clustered, component,
driving the accelerated expansion of the universe. This property requires a negative pressure. Many different models have been evoked to describe this dark sector of the energy content of
the universe, going from the inclusion of exotic components in the context of general relativity theory to modifications of the gravitational theory
itself, passing by other possibilities as the breakdown of the homogeneity condition. For a recent review, see reference \cite{caldwell}.
\par
A very appealing proposal to describe the dark sector are the so-called unified models. The protype of such model is the Chaplygin gas \cite{moschella,berto,jackiw}. In the unified
model dark matter and dark energy are described by a single fluid, which behaves as ordinary matter in the past, and as a cosmological constant
term in the future. In this sense, it interpolates the different periods of evolution of the universe, including the present stage of accelerated
expansion. The Chaplygin gas model leads to very good results when confronted with the obsevational data of supernova type Ia \cite{colistete}. Concerning the matter
power spectrum data, the statistic analysis leads to results competitive with the $\Lambda$CDM model, but the unified (called quartessence) scenario
must be imposed from the begining \cite{hermano1,hermano2}. This means that the only pressureless component admitted is the usual baryonic one,
otherwise there is a conflict between the contraints obtained from the matter power spectrum and the supernova tests. 
\par
Many variations of the Chaplygin gas model have been proposed in the literature. One of them is the Modified Chaplygin gas (MCG) Model. The equation of state of the MCGM is
\begin{equation}
\label{eos1}
p_c = B\rho - A\rho^{-\alpha},
\end{equation}
where $B$, $A$ and $\alpha$ are constants. When $B = 0$ we recover the Generalized Chaplygin Gas (GCG) Model, and if in addition
$\alpha = 1$ we have the original Chaplygin gas model.
The dynamics of the MCG model has been studied in the reference \cite{1}, while a dynamical system analysis has been made
in reference \cite{2}. The evolution of the temperature function has been considered in reference \cite{3}. Some
background constraints were established in references \cite{4} and \cite{5}. The analysis of the spherical collapse was made in reference \cite{6},
while a perturbative study, looking for some general features of the model, was carried out in reference \cite{7}. In all these studies the viability of the model was concluded, but no one of them
has exploited the observational data concerning the perturbative behaviour of the model.
\par
Our intention here is to test the MCG model against the power spectrum observational data. For the background tests, as those analyzed in
references \cite{4,5}, the MCG reveals to lead to competitive scenarios compared with the $\Lambda$CDM and the GCG models (to give just some
examples). However, the constraints coming from the power spectrum data are, in general, much more crucial since it tests not only the
background framework, but also the perturbative behaviour of the model. Many general constraints can be established on the parameters
of the EoS (\ref{eos1}) even not considering perturbations. For example, in the past the equation of state (\ref{eos1}) implies,
when $\alpha$ is positive (a requirement necessary in a perturbative analysis in order to preserve a positive sound speed) that
\begin{equation}
\rho_c(a\approx0)=\frac{cte}{a^{3(1+B)}}.
\end{equation}
In order not to spoil the usual primordial scenario of the standard model (in special nucleosynthesis), $B$ must be smaller than
$1/3$, which may include negative values. On the other hand, as we will show below, the requirement that the sound speed of the MCG
must be positive implies essentially that $B > 0$. Hence, the admissible values of the parameter $B$ seems to be around $ 0 < B < 1/3$. These considerations
will be strengthened through the power spectrum analysis to be made later in this paper: the matter power spectrum data can be fitted only if $|B| < 10^{-6}$. Hence, essentially the only configuration possible is that
corresponding to the generalized Chaplygin gas, with perhaps some possible very small deviations
from it. In this sense, we can consider that the
MCG model is ruled out when confronted with the power spectrum observational data.
\par
In next set we set out the general equations of the MCG model at background and perturbative levels. In section III we perform a numerical analysis
comparing the theoretical results with the matter power spectrum observational data. In section IV we present our conclusions.

\section{Basic set of equations}

Our starting point are Einstein's equations coupled to a pressureless fluid, radiation and to the MCG fluid. They read,
\begin{eqnarray}
R_{\mu\nu} &=& 8\pi G\biggr\{T^m_{\mu\nu} - \frac{1}{2}g_{\mu\nu}T^m\biggl\} + 8\pi G\biggr\{T^r_{\mu\nu} - \frac{1}{2}g_{\mu\nu}T^r\biggl\} + 8\pi G\biggr\{T^c_{\mu\nu} - \frac{1}{2}g_{\mu\nu}T^c\biggl\},\nonumber\\
{T_m^{\mu\nu}}_{;\mu} &=& 0 \quad , \quad {T_c^{\mu\nu}}_{;\mu} = 0 \quad , \quad {T_r^{\mu\nu}}_{;\mu} = 0 \nonumber
\end{eqnarray}
The superscripts (subscripts) $m$, $r$ and $c$ stand for "matter", "radiation" and "Chaplygin". We assume a perfect fluid structure for the cosmic medium as a whole and also for each of the components,
\begin{equation}
 \qquad T_{A}^{\mu\nu} = (\rho_{A} + p_A) u_A^{\mu} u^{\nu}_{A} - p_{A} g^{\mu\nu} \
,\qquad A = m, c, r\ .
\label{T}
\end{equation}
Using now the flat Friedmann-Robertson-Walker metric (as suggested by the Seven-year WMAP data \cite{komatsu}),
\begin{eqnarray}
ds^2 = dt^2 - a(t)^2[dx^2 + dy^2 + dz^2]\ ,\nonumber
\end{eqnarray}
and identifying all the background $4$-velocities,
Einstein's equations reduce to
\begin{eqnarray}
\biggr(\frac{\dot a}{a}\biggl)^2 &=& \frac{8\pi G}{3}\rho_m + \frac{8\pi G}{3}\rho_r + \frac{8\pi G}{3}\rho_c,\\
2\frac{\ddot a}{a} + \biggr(\frac{\dot a}{a}\biggl)^2 &=& - 8\pi G(p_c+p_r),\\
\dot\rho_m + 3\frac{\dot a}{a}\rho_m = 0 \quad &\Rightarrow& \quad \rho_m = \rho_{m0}/a^{3},\\
\dot\rho_r + 4\frac{\dot a}{a}\rho_r = 0 \quad &\Rightarrow& \quad \rho_r = \rho_{r0}/a^{4},\\
\dot\rho_c + 3\frac{\dot a}{a}(\rho_c + p_c) = 0 \quad (p_c =B\rho_c - A/\rho_c^\alpha) \quad &\Rightarrow& \quad \rho_c = \biggr\{A_s + \frac{1-A_s}{a^{3(1 + \alpha)(1+B)}}\biggl\}^{1/(1 + \alpha)}.
\end{eqnarray}
\par
In the above set of equations we have defined $A_s=\frac{A}{(1+B)\rho_{c0}^{1+\alpha}}$
\par
The perturbed equations in the synchronous coordinate condition can be established following closely the computation shown in reference
\cite{hermano2}. 
We introduce fluctuations around the background quantities, $g_{\mu\nu} = \bar g_{\mu\nu} + h_{\mu\nu}$, $\rho = \bar\rho + \delta\rho$,
$p = \bar p + \delta p$, $u^\mu = \bar u^\mu + \delta u^\mu$. The bars indicate the background quantities. The synchronous
coodinate condition implies $h_{\mu0} = 0$ and $\delta u^0 = 0$. The final perturbed equations read (see also \cite{weinberg}),
\begin{eqnarray}
\frac{\ddot{h}}{2}+\frac{\dot{a}}{a}\dot{h}-4\pi G\left(\delta \rho+3\,\delta p\right) &=& 0\\
\dot{\delta \rho}+\frac{3\dot{a}}{a}\left(\delta\rho+\delta\,p\right)+\left(\rho+p\right)\left(\theta-\frac{\dot{h}}{2}\right) &=& 0,\\
\left(p+\rho\right)\dot{\theta}+\left[\left(\dot{\rho}+\dot{p}\right)+\frac{5\dot{a}}{a}\left(\rho+p\right)\right]\theta+\frac{\nabla^{2}\delta\,p}{a^{2}} &=& 0,
\end{eqnarray}
where $\rho$ and $p$ stand for the total matter and pressure, respectively, $\theta = \delta u^{i}_{,i}$ and $h = h_{kk}/a^2$.
\par
In terms of the components, we end up with the following equations:
\begin{eqnarray}
\frac{\ddot{h}}{2}+\frac{\dot{a}}{a}\dot{h}-4\pi G\left[\delta \rho_{m}+\delta \rho_{c}+\delta \rho_{r}+3 (\delta \,p_{m}+\delta p_{c}+\delta \,p_{r})\right] &=& 0,\\
\dot{\delta \rho_{m}}+\frac{3\dot{a}}{a}\left(\delta\rho_{m}+\delta\,p_{m}\right)+\left(\rho_{m}+p_{m}\right)\left(\theta_{m}-\frac{\dot{h}}{2}\right) &=& 0,\\
\left(\rho_{m}+p_{m}\right)\dot{\theta_{m}}+\left[\left(\dot{\rho_{m}}+\dot{p_{m}}\right)+\frac{5\dot{a}}{a}\left(\rho_{m}+p_{m}\right)\right]\theta_{m}+\frac{\nabla^{2}\delta\,p_{m}}{a^{2}} &=& 0,\\
\dot{\delta \rho_{c}}+\frac{3\dot{a}}{a}\left(\delta\rho_{c}+\delta\,p_{c}\right)+\left(\rho_{c}+p_{c}\right)\left(\theta_{c}-\frac{\dot{h}}{2}\right) &=& 0,\\
\left(\rho_{c}+p_{c}\right)\dot{\theta_{c}}+\left[\left(\dot{\rho_{c}}+\dot{p_{c}}\right)+\frac{5\dot{a}}{a}\left(\rho_{c}+p_{c}\right)\right]\theta_{c}+\frac{\nabla^{2}\delta\,p_{c}}{a^{2}} &=& 0,\\
\dot{\delta \rho_{r}}+\frac{3\dot{a}}{a}\left(\delta\rho_{r}+\delta\,p_{r}\right)+\left(\rho_{r}+p_{r}\right)\left(\theta_{r}-\frac{\dot{h}}{2}\right) &=& 0,\\
\left(\rho_{r}+p_{r}\right)\dot{\theta_{r}}+\left[\left(\dot{\rho_{r}}+\dot{p_{r}}\right)+\frac{5\dot{a}}{a}\left(\rho_{r}+p_{r}\right)\right]\theta_{r}+\frac{\nabla^{2}\delta\,p_{r}}{a^{2}} &=& 0 \ ,
\end{eqnarray}
with $\theta_{m} = \delta u_{m,i}^{i}$, $\theta_{c} = \delta u_{c,i}^{i}$ and $\theta_{r} = \delta u_{r,i}^{i}$.
\par
With the definitions
\begin{eqnarray}
\Omega_c(a) &=& \Omega_{c0}\left(A_s+\frac{1-A_s}{a^{3\left(1+\alpha\right)\left(1+B\right)}}\right)^{\frac{1}{1+\alpha}},\\
w(a)&=&\frac{p_c}{\rho_c}\hspace{0.1cm}=\hspace{0.1cm} B -\frac{ A_s(1+B)}{A_s+(1-A_s)a^{-3(1+\alpha)(1+B)}},\\
v^{2}_{s}(a) &=& B+\frac{\alpha A_s(1+B)}{A_s+(1-A_s)a^{-3(1+\alpha)(1+B)}},\\
H(a)&=& \left(\frac{\Omega_{m0}}{a^{3}}+\Omega_{c}\left(a\right)+\frac{\Omega_{r0}}{a^{4}}\right)^{1/2},\\
q(a)&=&\frac{\frac{\Omega_{m0}}{a^{3}}+\Omega_{c}\left(a\right)(1+3w(a))+\frac{2\Omega_{r0}}{a^{4}}}{2\left(\frac{\Omega_{m0}}{a^{3}}+\Omega_{c}\left(a\right)+\frac{\Omega_{r0}}{a^{4}}\right)}
\end{eqnarray}
and remembering that $p_m = \delta p_m = 0$, leading also to $\theta_m = 0$ up to an irrelevant non-homogeneous term,
the set equations becomes
\begin{eqnarray}
\delta^{\prime\prime}+\left[2-q\left(a\right)\right]\frac{\delta^{\prime}}{a}
-\frac{3\Omega_{m0}}{2a^{5}\left[H(a)\right]^{2}}\delta &=& \frac{3\Omega_{c}(a)}{2\left[aH(a)\right]^{2}}\lambda\left[1+3v_{s}^{2}(a)\right]+\frac{3\Omega_{r0}}{a^{6}\left[H(a)\right]^{2}}\delta_{r},
\\
\lambda^{\prime}+\frac{3}{a}\left[v_{s}\left(a\right)-w\left(a\right)\right]\lambda &=& - \left[1+w\left(a\right)\right]
\left[\frac{\theta_{c}(a)}{aH(a)}-\delta^{\prime}\right],\\
\left[1+w\left(a\right)\right]\left\{\theta_{c}^{\prime}+\frac{\left[2-3v_{s}^{2}(a)\right]}{a}\theta_{c}\right\} &=&v^{2}_{s}(a)\left(\frac{k}{k_{0}}\right)^{2}\frac{\lambda}{H(a)a^{3}},\\
\delta_{r}^{\prime}+\frac{4}{3}\left(\frac{\theta_{r}}{aH(a)}-\delta^{\prime}\right) &=& 0,\\
\theta_{r}^{\prime}+\frac{\theta_{r}(a)}{a} &=& \left(\frac{k}{k_{0}}\right)^{2}\frac{\delta_{r}}{4H(a)a^{3}}\ ,
\end{eqnarray}
where
\begin{equation}
\delta \equiv \frac{\delta\rho_{m}}{\rho_{m}}\ , \quad \lambda \equiv \frac{\delta\rho_{c}}{\rho_{c}}\ ,\quad\delta_{r} \equiv \frac{\delta\rho_{r}}{\rho_{r}}
\
\label{}
\end{equation}
and $k_{0}^{-1}=3000\,h\,Mpc$. A Fourier decomposition of the spatial dependence of the perturbed quantities has been performed, $k$ being
the corresponding Fourier mode.

\section{Numerical analysis}

The matter power spectrum is defined by
\begin{equation}
{\cal P} = \delta_k^2 \quad ,
\end{equation}
where $\delta_k$ is the Fourier transform of the dimensionless density
contrast $\delta_m$. In what follows we will use the matter power spectrum data of the 2dFGRS observational mapping \cite{2d}. We use
the data in the range $ 0.01 Mpc^{-1}h < k < 0.185 Mpc^{-1}h$ (here $h$ is connected with the Hubble parameter today being defined by $H_0 = 100\,h\,km/Mpc.s$,
with no relation to the previous definition of the metric fluctuation) which corresponds to the linear regime which is being considered here. Hence,
the covariance matrix is diagonal.
\par
One important aspect of the numerical analysis is to fix the initial conditions. In order to do so, we use a scale invariant primordial
spectrum, with the BBKS transfer function \cite{bardeen}. The initial conditions are fixed following in general lines the prescription
described in reference \cite{sola}.
Even if a full statistical analysis may be performed using, for example, a bayesian statistical, such complete analysis is not essential to obtain
the main result of this work. 
\par
In any  perturbative analysis, one crucial quantity is the squared sound speed. Positive values are assured
if $B$ and $\alpha$ are positive, see equation (21). It is possible to have also positive squared speed in the case $\alpha$ is negative, but in a very small range.
The possibility to have $v_s^2 > 0$ and $B < 0$ is almost excluded, as it can be seen in figure 1.
However, in performing the numerical computation we will consider this possibility.
\par
In figure 2 we plot the linear matter power spectrum comparing with the observational data for $A_s = 0.95$ and $\alpha = 10$ (left pannel) and $\alpha = 1$ (center pannel), with different values of $B$. In the right pannel the two-dimensional Probability
Distribution Function is displayed, showing the contours at $1\sigma$, $2\sigma$ and
$3\sigma$ confidence levels. When $\alpha = 10$, the case $B = 0$ is essentially the only one that fits
the observational data, with a total $\chi^2$ of about $17$ (the statistical parameter that measure, to say in general lines, the quality of the
fitting of the data by a theoretical model, the smaller the value of $\chi^2$ the better the fitting). As it can be verified, when $|B| > 10^{-4}$ there is already a huge discrepancy between the theoretical results and
the observational data. At the center pannel of the same figure, the only changing is $\alpha = 1$. Now the better fitting is achieved
by $B = - 2.7\times10^{-4}$, but with $\chi^2 \sim 37$: the fitting is much worse than in preceding case. An extensive
inspection of the different possibilities shows that a reasonable agreement, at $1\sigma$ confidence level, can be find only around $|B| < 10^{-6}$, with $\alpha >> 1$ or
$\alpha \approx 0$. Such fine tuning in the EoS parameter (\ref{eos1}) 
implies that the only "natural"' value would be $B = 0$. This reduces the model to the the GCG model. Hence, the confrontation with the matter
power spectrum data seems to rule out the MCG model.

\begin{center}
\begin{figure}[!t]
\begin{minipage}[t]{0.3\linewidth}
\includegraphics[width=\linewidth]{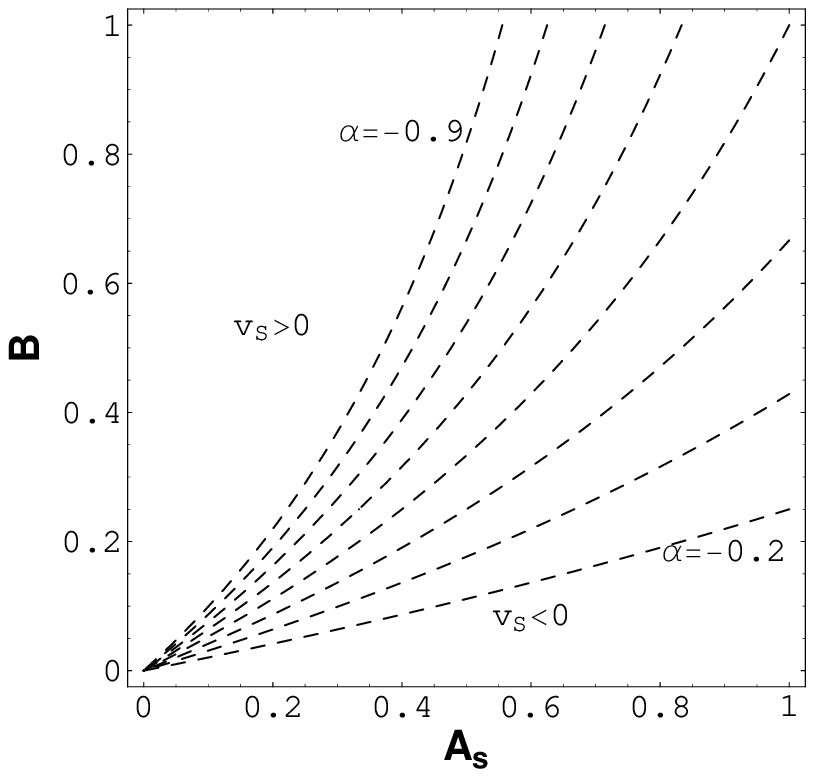}
\end{minipage} \hfill
\begin{minipage}[t]{0.3\linewidth}
\includegraphics[width=\linewidth]{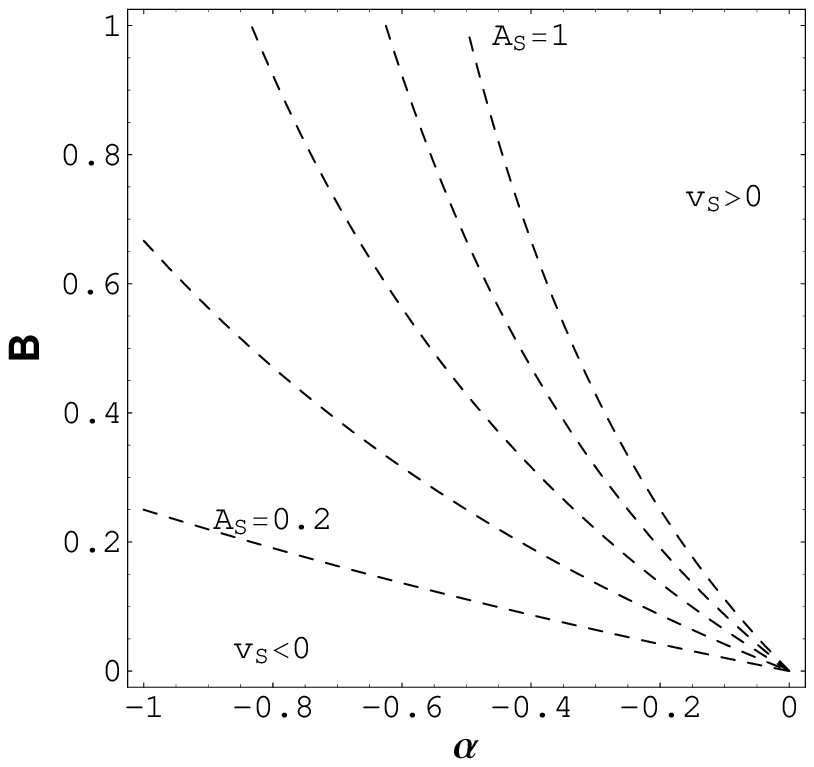}
\end{minipage} \hfill
\begin{minipage}[t]{0.3\linewidth}
\includegraphics[width=\linewidth]{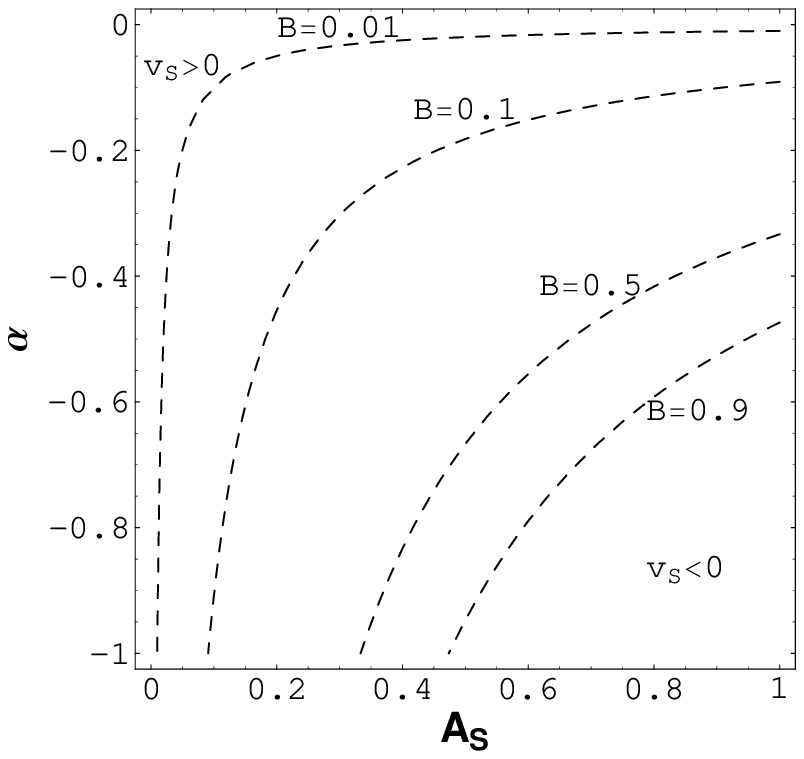}
\end{minipage} \hfill
\caption{{\protect\footnotesize In each panel we have fixed one of the parameters ($\alpha$,$A_s$,$B$) and ploted the contours for which the speed of sound is equal to zero. Above (below) each dashed line we have $v^{2}_{s}>0\, (v^{2}_{s}<0)$ for differents values of $\alpha-(A_s)-(B)$ in the left-(center)-(right) panel. }}
\label{Fig1}
\end{figure}
\end{center}

\begin{center}
\begin{figure}[!t]
\begin{minipage}[t]{0.3\linewidth}
\includegraphics[width=\linewidth]{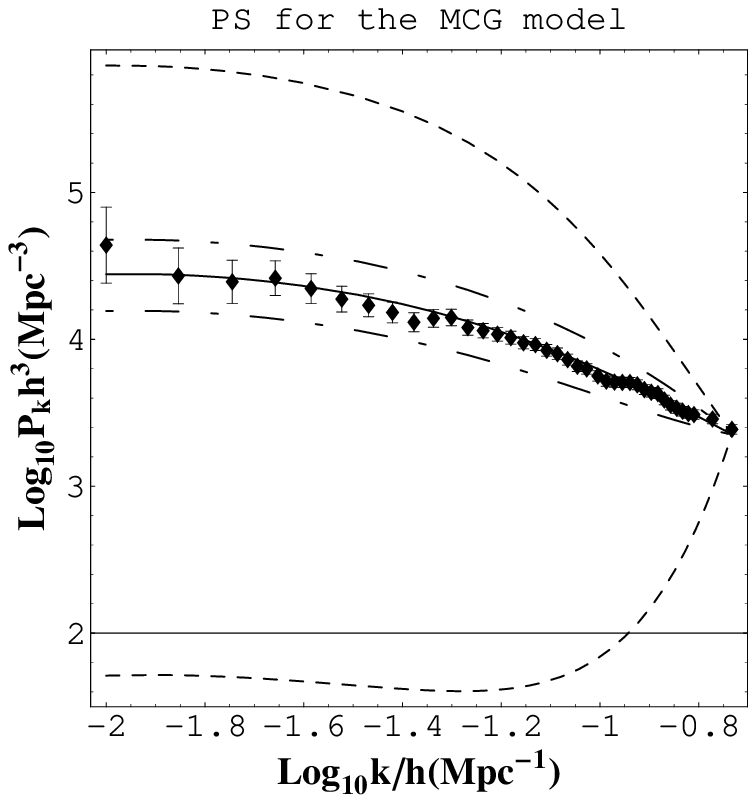}
\end{minipage} \hfill
\begin{minipage}[t]{0.3\linewidth}
\includegraphics[width=\linewidth]{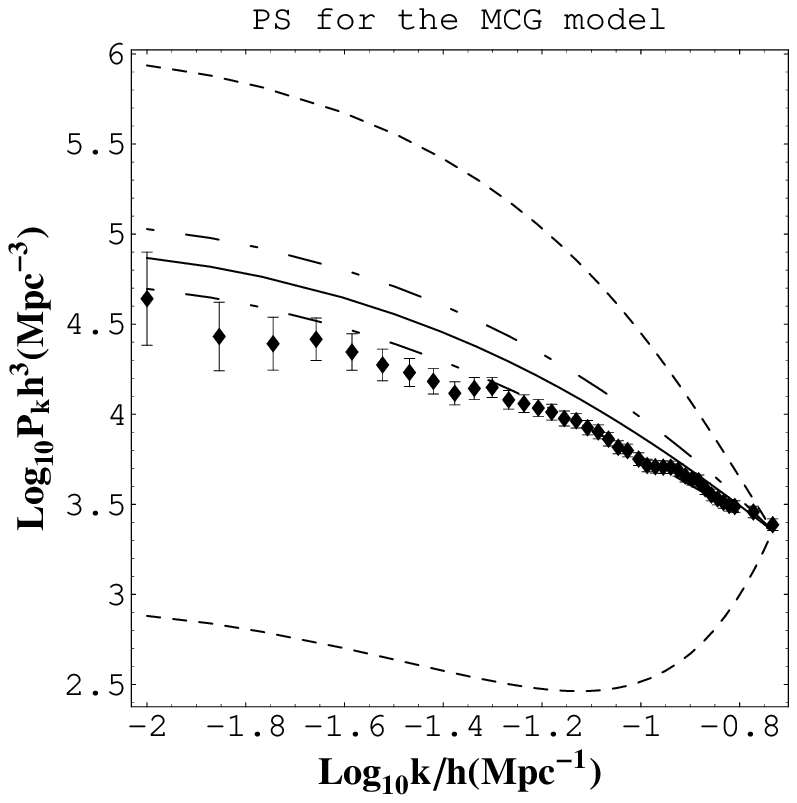}
\end{minipage} \hfill
\begin{minipage}[t]{0.35\linewidth}
\includegraphics[width=\linewidth]{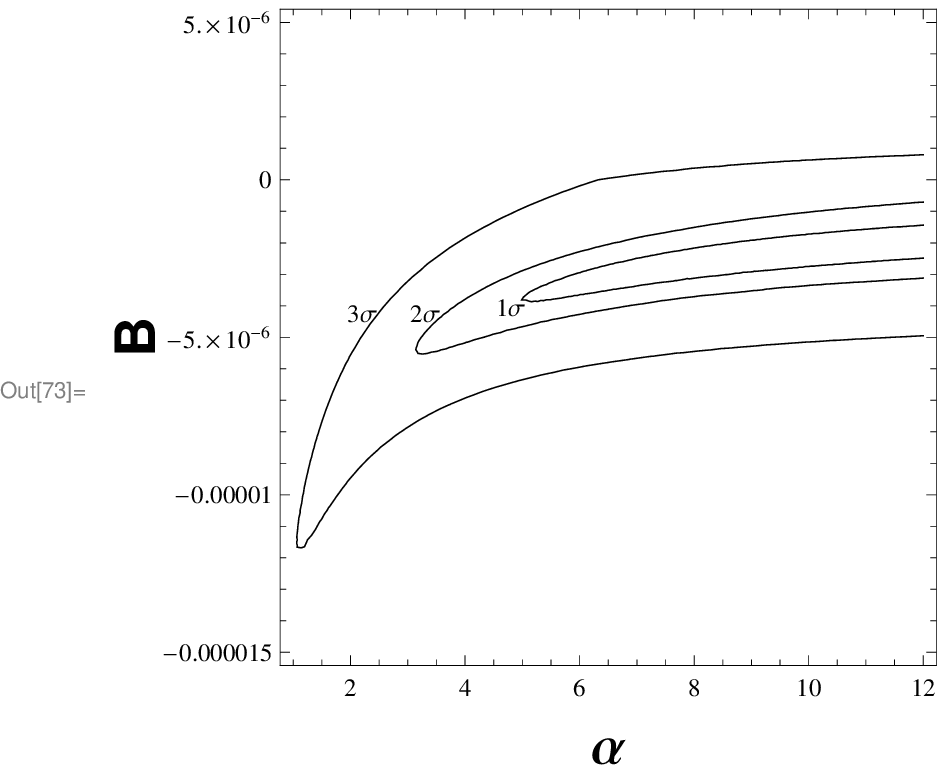}
\end{minipage} \hfill
\caption{{\protect\footnotesize At left, power spectrum for the MCG model fixing $A_s=0.95$ and $\alpha=10$. From top to bottom, the curves are MCG models with
$B=10^{-4}$, $B=10^{-5}$, $B=0$, $B=-10^{-5}$ and $B=-10^{-4}$, respectively  At center, power spectrum for $\alpha=1.00$, and for the same values
as before of the parameters $A_s$ and $B$. In the right pannel it is displayed the $1\sigma$, $2\sigma$ and $3\sigma$ confidence levels contours for the
two-dimensional Probability
Distribution Function for $B$ and $\alpha$. }}
\label{Fig2}
\end{figure}
\end{center}

\section{Conclusions}

The Modified Chaplygin Gas (MCG) model is an extension of the Generalized Chaplygin Gas (GCG) model, with the addition of a new term in the usual
GCG equation of state which is proportional to the density. In this sense, it can be seen as a combination of the CGM model (where
$p = - A/\rho_c^\alpha$) and the so-called
XCDM model (where $p = B\rho$). In the present paper we have confronted the MCG model against the power spectrum observational data. In general,
the MCG leads to good behaviour concerning the background tests. However, the power spectrum analysis implies to perform a perturbative study
of the model, testing in this sense the deep content of the theoretical framework. In this perturbative analysis, a crucial quantity is
the squared sound speed which must be positive in order to avoid instabilities. In the MCG model, the positivity of the squared sound speed
requires $B$ and $\alpha$ to be also positive, even if small range of negative values of $B$ and $\alpha$ can in principle be admitted.
\par
However, the confrontation of the MCG model with the power spectrum observational data rules out any significant departures from $B = 0$. In fact,
only values such that $|B| < 10^{-6}$the model can fit the observational data. This result essentially rule out the MCG model, reducing it to the
GCG model. Crossing these results with those of reference \cite{hermano1,hermano2}, where power spectrum constraints on the GCG model were establisheds, only the quartessence version of the GCG model survives. In the quartessence formulation of the GCG model, the matter content
is
given only by the baryonic component. If this quantity is left free, we find a prediction of a universe dominated only by matter, in
contradiction to the supernova analysis \cite{colistete}.
\par
It must be remarked however that the conclusions obtained in the present work are restricted to a hydrodynamical representation of the
dark energy component. If this component is alternatively represented by a self-interacting scalar field, leading to the same background
relations, the restrictions arising from the positivity of the squared sound speed disappears. If the MCG model can survive the observational
tests in this
alternative formulation is a question we intend to address in a future work. The main difficult comes from the fact that the usual
GCG gas is connected a kind of Born-Infeld , which leads to essentially the same expression for the sound speed as in the hydrodynamical formulation \cite{morita}. 
Hence the restrictions found here for the hydrodynamical formulation may remain. If such problem can be circumvented in the case MCG model is an open question.
\newline
\noindent
{\bf Acknowledgements:} J.C.F. and H.E.S.V. thank CNPq (Brazil) and FAPES (Brazil) for partial financial support. C.O. and J.T. thank the
Africa-Brazil scientific cooperation program {\it Pr\'o-Africa - CNPq} for partial financial support, and the Departamento
de F\'{\i}sica of UFES for the kind hospitality during part of the elaboration of this work.

\end{document}